\input{aipcheck}
\documentclass[numberedheadings,final]{aipproc}
\usepackage{natbib}
\usepackage{amssymb}
\usepackage{amsmath}

\layoutstyle{8x11single}

\newcommand{\Mpch}{{$h^{-1}$Mpc}}
\newcommand{\kms}{{km.s$^{-1}$}}
\newcommand{\kmsMpc}{{km.s$^{-1}$.Mpc$^{-1}$}}

\begin{document}

\title{Dynamics of the Local Universe:\\ cosmic velocity flows and voids}

\author{Guilhem Lavaux}{
  address={University of Illinois at Urbana-Champaign, 1110 W Green St, 61801 Urbana, IL, USA}
}

\begin{abstract}
A valuable amount of information is available in peculiar velocities
of galaxies. Peculiar velocity surveys have recently allowed the discovery
of potential problems with $\Lambda$CDM \citep[e.g][]{Watkins09}. Nonetheless,
their direct observation through distance measurements remains a daunting task.
Another way of considering the problem is to use orbit reconstruction methods 
assuming some mass-to-light assignment for galaxies. We give here two applications
of this procedure for the study of large-scale bulk flows and the dynamics of
voids in the Local Universe. We concentrate our study on the use of the
Monge-Amp\`ere-Kantorovitch reconstruction technique.

Using peculiar velocities reconstructed from the 2MASS Galaxy
Redshift Survey, and after comparison with the NGB-3k peculiar
velocity catalog, we look in the details of these peculiar
velocities. More particularly, we estimate the constraints that
the peculiar velocities put on the cosmology.

The information on dynamics that is included in reconstructed orbits
of galaxies also allows us to have a much better prescription for
defining and identifying voids in simulations and redshift
catalogs. We present this new technique and how voids
may give us additional constraints on cosmology with current and future surveys.
\end{abstract}

\maketitle

\section{Introduction}

The dynamical study of large-scale structures holds an important wealth
of information on cosmology. So far, most of the effort has been put
on inferring and interpreting from galaxies the current matter density
field. While this is a fair approach as it is easier to consider in a
first time, it is also more complicated because of non-linear effects
between the primordial fluctuations and the present observed matter
fluctuations.

Now tools exist to explore much more linear quantities like
peculiar velocities and the displacement field. These two components
are related to the density field, while being
mostly linear in the primordial fluctuations. This linearity allows us to make
much quicker predictions and comparisons with different cosmological
models. This remark is particularly true for the study of voids, whose
exploitation as a probe for cosmological parameters is hindered by the
complexity of the analysis of the topography of the density
field. Additionally, studying peculiar velocities allows us to make a
consistency check with larger peculiar velocity measurements that are
now appearing.

Another major interest of studying peculiar velocities is their higher sensitivity
to larger scales compared to the density field, which is more sensitive to smaller scales. By comparison to observational
data, these velocities allow us to probe deeper, or hidden part, of the Universe by studying
their local amplitudes and directions.

This proceeding is organized as follow. In
Section~\ref{sec:veloc_flows}, we discuss the use of peculiar
velocities as a direct probe of the Universe mean matter density $\Omega_\text{m}$ and what are the
last results obtained using a non-linear prescription for the
reconstruction of peculiar velocities. In Section~\ref{sec:void_dyn},
we discuss other ways to use the reconstruction method used in
peculiar velocity analysis (Section~\ref{sec:rec_veloc}
and \ref{sec:mak}) to study the void dynamics through the statistics
of the displacement field. In Section~\ref{sec:conclu}, we conclude.

\section{Cosmic velocity flows}
\label{sec:veloc_flows}

The peculiar velocities, being directly sensitive to the total matter content, are more and more studied with the advent of large luminosity distance catalogs \citep[e.g.][]{SFI1,SFI2,SFI++,2MTF}. However the interpretation  of peculiar velocities is rendered difficult by two major problems: the statistics is complicated and the volume of the catalogs is relatively small compared to the current galaxy redshift survey standards. 
We look into recent work that tries to use non-linear methods of reconstruction to predict the peculiar velocity field and then compare it to observations.

We recall in Section~\ref{sec:obs_veloc} how to observe directly peculiar velocities and why we need the reconstruction procedure in Section~\ref{sec:rec_veloc}. We describe the Monge-Amp\`ere-Kantorovitch (MAK) reconstruction in Section~\ref{sec:mak} and its performance at reconstructing peculiar velocity field in Section~\ref{sec:mak_perf}. We then look at the particular peculiar velocities of our Local Universe in Section~\ref{sec:pec_locuniv}.

\subsection{Observing peculiar velocities}
\label{sec:obs_veloc}

While peculiar velocities hold valuable information on the dynamics,
they are very difficult to observe. Currently, there is just one way
to obtain them on large scales, which is a few tenth of
Megaparsecs. First, one compute the redshift $z$ of a galaxy using
Doppler effect on the galaxy spectrum. Second, one must have an
indicator of the Luminosity distance, $d$. This distance is
independent of redshift measurement and the indicator must only rely
on distance independent physics. Several indicators exist nowadays:
\begin{itemize}
\item[-] Primary indicators, like the Cepheid luminosity-period
  relation gives the distance of the nearest galaxies where it is
  possible to resolve these stars. A relatively recent technique, which qualifies as
  primary, makes use of the Water maser
  emission in blobs around some of the galaxies \citep{Herrnstein99}.
\item[-] Then we must rely on secondary precise indicators. First for
  slightly farther away galaxies we may use the Tip of the Red Giant
  Branch, when stars are resolved \citep{tip_giant1,tip_giant2}, or Surface Brightness
  Fluctuations, when they are just at the limit of being resolved
  \citep{sbf1,sbf2}.
\item[-] For much farther galaxies, we have to rely on less precise
  indicators like Tully-Fisher relation \citep{TF77}, for spiral galaxies, or
  Faber-Jackson relation \citep{FJ76}, for elliptical galaxies.
\end{itemize}
These two distances may be linked together to form the Hubble relation
\begin{equation}
        c z = H d + v_r
\end{equation}
with $H$ the Hubble constant, $v_r$ the line of sight component of the
peculiar velocity of the considered galaxy, $c$ the speed of light.

Three problems quickly arise. First, we have only one component of the
3d peculiar velocity. Second, $v_r$ is very noisy as it is itself a
perturbation on the distance-distance relation. Third, it is time
consuming to obtain the distance $d$ from observations so we only have
access to the distances of a few hundreds of galaxies.

\subsection{Reconstructing peculiar velocities}
\label{sec:rec_veloc}

The aforementioned problems of measuring peculiar velocities have long hindered their use as a
way to study dynamics. We propose another way to study this problem
through methods of reconstruction of peculiar velocities. While
methods have existed for a few decades now \citep[e.g][]{Peebles76}, they were all mostly based on
applying linear theory. We investigated a new reconstruction scheme
called the Monge-Amp\`ere-Kantorovitch reconstruction
\citep{Brenier2002,Lavaux08}.

Methods of reconstruction are all based on the following general
scheme. We first start from a redshift catalog of galaxies, from which
we obtain the redshift positions and an estimate of the intrinsic
luminosity. Assuming that light traces mass, we derive a mass
catalog. This catalog is used by a reconstruction algorithm to predict
peculiar velocities.

It is then possible to derive cosmological parameters by considering
the subset of objects for which peculiar velocities were observed, and
thus comparing the reconstructed velocities to the observed
velocities. From another point of view, studying the reconstructed
velocities itself is interesting to study the origin of the CMB
dipole, the dynamics of voids and, e.g., the initial conditions of our
Local Universe.

\subsection{The Monge-Amp\`ere-Kantorovitch reconstruction (MAK)}
\label{sec:mak}

The MAK reconstruction is based on an observation that first order
Lagrangian perturbative theory is very effective at even describing
non-linear effects in Eulerian theory. This motivated the assumption
that the mapping between Lagrangian coordinates ${\bf q}$ and present,
Eulerian, coordinates ${\bf x}$ must be essentially convex and
potential.\footnote{A potential vector field satisfies the condition that
there exists a scalar field from which this vector field is the gradient.
The convexity enforces that this potential is convex at any spatial point.}
 This is equivalent to assuming that shell-crossed region
represents a negligible volume of the whole Universe. Adding to this
assumption the continuity equation, we obtain the Monge-Amp\`ere
equation:
\begin{equation}
  \text{det}\left( \frac{\partial^2 \xi}{\partial q_i \partial
    q_j}\right) = \frac{\rho({\bf x}({\bf
      q}),t)}{\rho_0} \label{eq:monge}
\end{equation}  
with $\rho({\bf x},t)$ the mass density field at the time $t$,
$\rho_0$ the original mass density field taken as constant over the
Universe and $\xi({\bf q})$ is the potential defining the mapping ${\bf q}\rightarrow {\bf x}$ 
by the relation ${\bf x}({\bf q}) = \nabla_{q} \xi$. \cite{Brenier2002} showed that solving Eq.~\eqref{eq:monge}
is equivalent to solving a Monge-Kantorovitch problem, which
corresponds to minimize the quantity
\begin{equation}
  I[{\bf q}({\bf x})] = \int_{\mathfrak{R}^3} \rho({\bf x}) |{\bf
    x}-{\bf q}({\bf x})|^2\,\,\text{d}^3{\bf x} \label{eq:kanto}
\end{equation}
according to the mapping ${\bf q}({\bf x})$. Discretizing this
equation into equal mass particles leads to a seemingly simple
quantity to minimize:
\begin{equation}
  S_\sigma = m_0 \sum_{i=1}^N \left({\bf x}_i - {\bf
    q}_{\sigma(i)}\right)^2
\end{equation}
according to the pairing maps $\sigma$, where ${\bf q}_i$ are
homogeneously distributed, ${\bf x}_i$ are distributed according to
$\rho({\bf x})$. The minimization algorithm is in itself sophisticated
as the direct approach would have a complexity of $O(N!)$, with $N$
the number of particles in the gravitational system. We use the
Auction algorithm \citep{Bertsekas79,Bertsekas1998,Brenier2002} to
solve this problem, which has a complexity of the order of $O(n^{2.25})$,
with $n$ the mean spatial density of particles.

\subsection{Performance of the reconstruction on mock samples}
\label{sec:mak_perf}

\begin{figure}
  \includegraphics[width=\hsize]{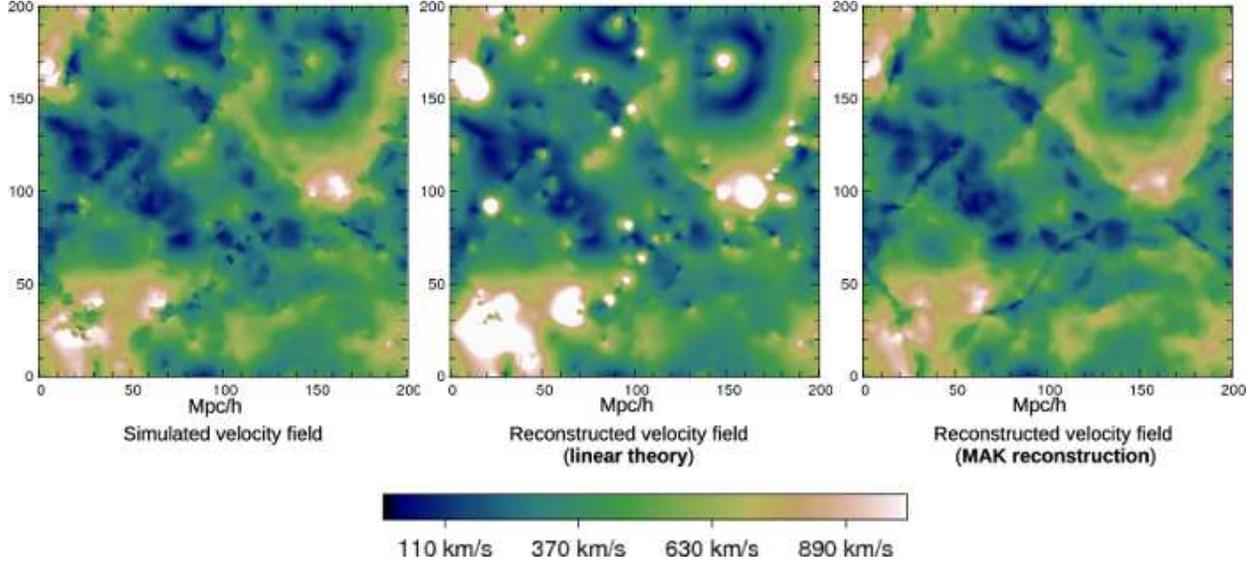}
  \caption{\label{fig:mak_simu} {\it Illustration of the performance of reconstruction algorithms on a $N$-body simulation} -- Left panel: velocity field as given by the $N$-body simulation after evolution from Gaussian initial conditions. Middle panel: velocity field reconstructed using linear theory and the final dark matter distribution of the simulation. Right panel: velocity field reconstructed using the Monge-Amp\`ere-Kantorovitch method and the final dark matter distribution of the simulation. The amplitude of the velocity field in a thin slice at the center of the simulation volume is represented in each panel. The scale is given by the bar below the panels.}
\end{figure}

We give in Fig.~\ref{fig:mak_simu} an illustration of the performance of
the reconstruction algorithms, linear theory and MAK, on a $N$-body
simulation. We obtained the simulated velocity field on the left panel
by running a $N$-body code on randomly generated initial
conditions and adaptively smoothing the velocities obtained at the end
of the run. We assumed the following cosmology for running the simulation: $\Omega_\text{m}=0.30$, $\Omega_\Lambda=0.70$, $H=65$~\kmsMpc, $n_\text{S}=1$, $\sigma_8=1.0$, $\Omega_\text{b}=0$. The $N$-body sample itself is composed of 128$^3$
particles in a box with a 200\Mpch{} side. The middle panel is
obtained by computing the gravitational field of the particles of the
simulation at their position at the end of the run. The right panel is
obtained by applying the MAK
reconstruction on the position of the particles at the end of the
simulation. The adaptive smoothing is effectively using a scale of a few \Mpch{}.

While the comparison between the middle and the left panel is fair,
the resemblance between the right panel and the left panel is
striking. Compared to linear theory, we nearly recover all the
features of the peculiar velocity field. The low velocity region are
perfectly reconstructed while the high velocity region suffers a few
problems. For example, the high velocity peak located at about
(150,175)\Mpch{} in the left panel is smeared out in the right
panel. But it is still a lot better than the huge peak that is
present in the middle panel at the same position.

While $N$-body sample gives an expected fair representation of the
large scale dynamics, galaxy redshift catalog are unfortunately not
that precise and complete due to intrinsic observational limitations.
\cite{Lavaux08} studied in details a lot of these limitations. We
present in Table~\ref{tab:obs_limit} a summary of the impact of these
problems on the actual measurement of $\Omega_\text{m}$ obtained by
looking at the relation between reconstructed and observed
velocities. We note that we are essentially limited by the effect of
the bias, which is our prior on the mass distribution according to the distribution
of galaxies, for the determination of
$\Omega_\text{m}$ from the velocity-velocity comparison. But, this
does not increase scatter and the impact of the other effects seems
under control.

\begin{table}
  \begin{tabular}{ccl}
    \hline
    Observational problem class & Effect & Observed systematic on $\Omega_\text{m}$ \\
    \hline
    Mass distribution & Incompleteness & 3-8\% \\
     & M/L assignment & 20-46\% \\
     & Diffuse mass & 50\% \\
    \\
    Redshift distortions & Finger-of-god & Small impact \\
     & Kaiser effect & no large scale systematic\\
    & & potential small scale effect \\
    \\
    Edge and finite volume effects & Zone of avoidance & No systematic, boundaries are affected \\
     & Tidal effects & No systematic \\
     & Cosmic variance & $\sim$20\% \\
    \\
    Statistical comparison & Velocity field correlation & Significant underestimation \\
    & & of error bar \\
    \hline
  \end{tabular}
  \caption{\label{tab:obs_limit} Observational effects on the measure of $\Omega_\text{m}$ from velocity-velocity comparison in redshift catalogs}
\end{table}

\subsection{The peculiar velocities in the Local Universe}
\label{sec:pec_locuniv}

As now we have assessed the importance of observational effects, we
may reconstruct the peculiar velocities of our Local Universe. To
achieve that goal, we used two catalogs: the Two-Micron All Sky
Redshift Survey (2MRS) and the Nearby Galaxy 3,000~\kms{} (NBG-3k)
distance catalog.

The 2MRS is based on the Two-Micron All Sky Survey (2MASS) photometric
galaxy catalog. It is composed of about 25,000 galaxies, selected in
the $K_\text{S}$ band with a magnitude limit of 11.25. Selecting according to $K_\text{S}$
allows us to have a fair sampling of the stellar mass in the 2MRS catalog by being sensitive mostly to the old population stars. It was then hoped to be a better indicator of the total galactic mass. \cite{L04} and \cite{Ramella04} showed that the luminosity in the K band is indeed a relatively good tracer of the mass with a mild dependence of the mass-to-light ratio to the luminosity.
 The other main
advantage of the 2MRS is to be full sky and complete down to a galactic latitude
of $b\sim 5-10$~degrees, depending whether we look in the direction
of the galactic bulge. So, we are limited by the strong obscuration
of galaxies in the direction of Milky Way's galactic plane. The
redshift galaxy distribution peaks at about $\sim 60-90$\Mpch{}. This
distribution is falling quickly above $\sim 120$\Mpch{}, with only a
handful of galaxies above $\sim 250$\Mpch{}.

The NBG-3k is a 30-40\Mpch{} deep galaxy catalog
\citep{TullyVoid07}. It contains 1791 galaxies with high quality
distances obtained from four different methods, when possible: the
Tully-Fisher relation, the Tip of the red giant branch, the Surface
Brightness fluctuation and the Fundamental plane.

\begin{figure}
  \includegraphics[width=\hsize]{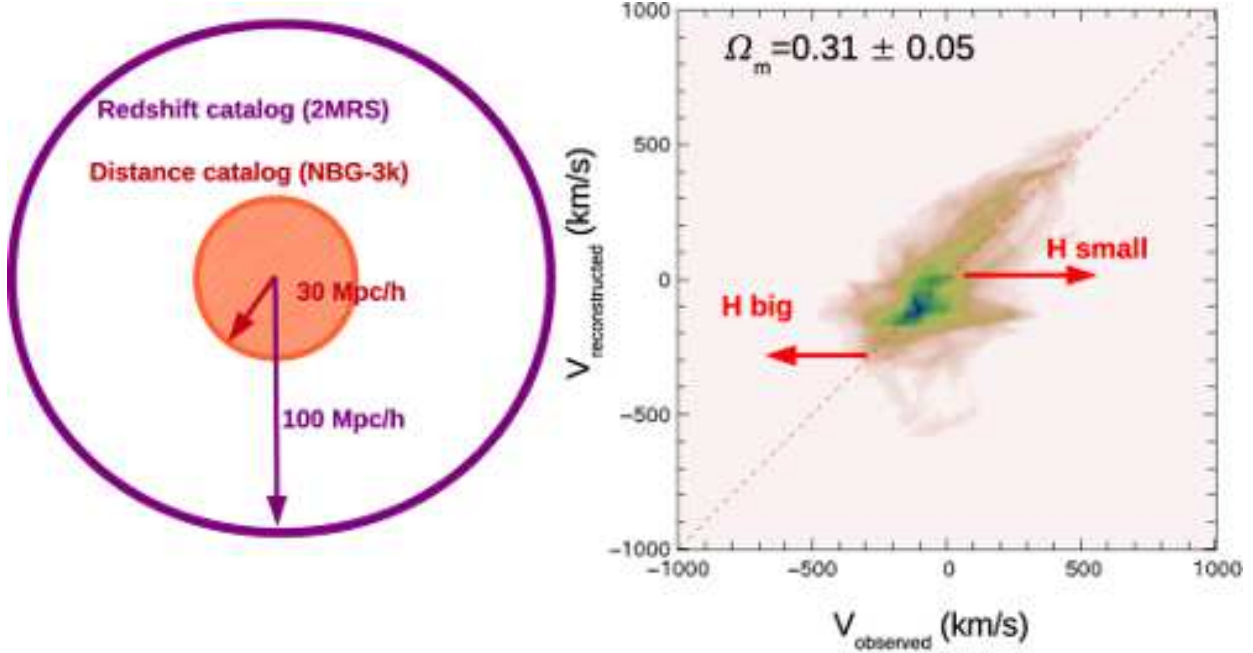}
  \caption{\label{fig:omega} {\it $\Omega_\text{M}$ estimation from peculiar velocities} -- Left panel: illustration giving the method of reconstruction and comparison. The reconstruction is achieved with a 100\Mpch{} deep survey and the comparison is done on a 30\Mpch{} deep volume where the two surveys overlap. Right panel: Scatter giving the relation between the reconstructed velocity against the observed velocity. The slope is relatively insensitive to the choice of the Hubble constant, which acts mostly by shifting the scatter along the horizontal axis. Here we assumed $H=83$~\kmsMpc{} and $\Omega_\text{m}=0.31$ for representing the scatter.}
\end{figure}

Using the 2MRS, we run a 100\Mpch{} deep reconstruction for a
comparison with observed velocities in a 30\Mpch{} deep volume. This
is sufficiently big to mitigate boundary effects in the volume of
comparison, while not too big to be strongly affected by shot noise
effects due to poorer mass sampling above 60\Mpch{}. More details on this
reconstruction are given in \cite{lavaux09}.

In Fig.~\ref{fig:omega}, we represented the result of the comparison
between reconstructed and observed velocities in the 30\Mpch{} deep
volume. We estimated $\Omega_\text{m} = 0.31\pm 0.05$ using the slope
of the scatter plot \citep{Lavaux08}. This slope is independent of any
arbitrary calibration problem in the Hubble constant we determined
from data. The Hubble constant would move the scatter along the
horizontal axis but would not change the slope.

\begin{figure}
  \includegraphics[width=\hsize]{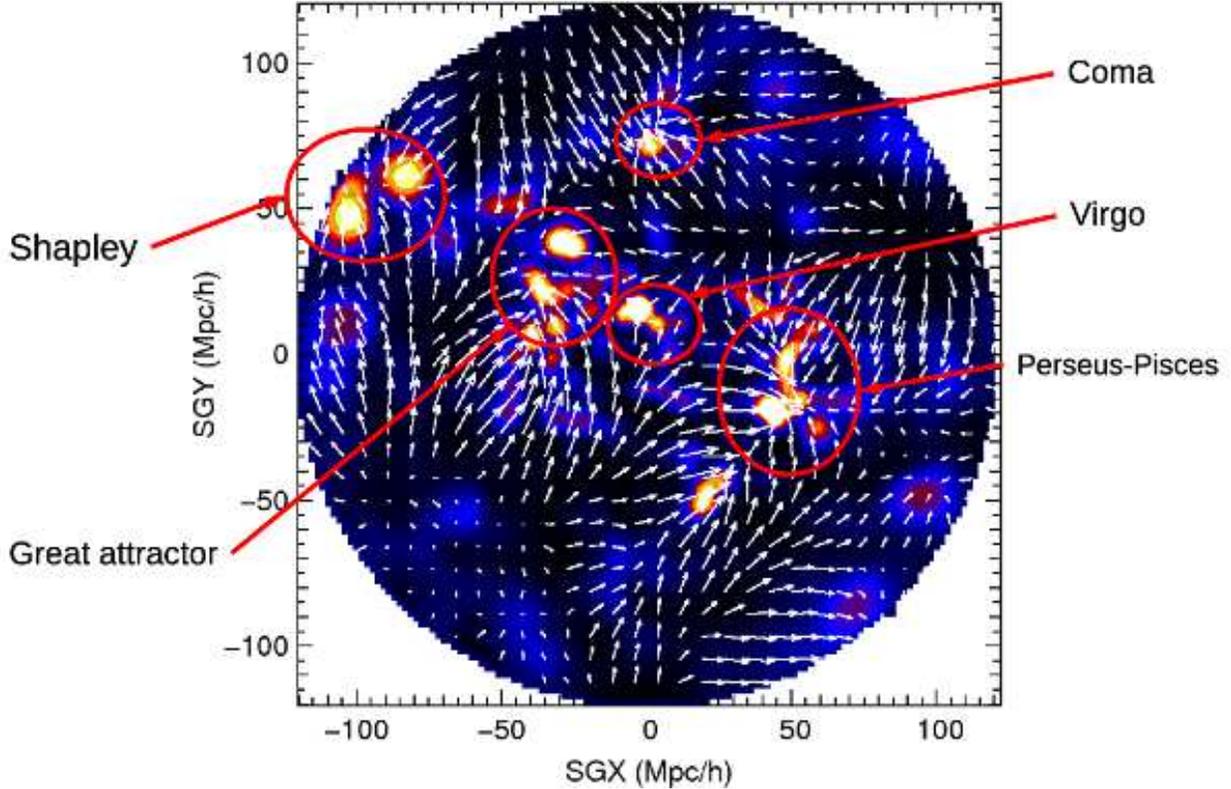}
  \caption{\label{fig:flows} {\it Reconstructed velocity field in the Supergalactic plane} -- We represent the reconstructed velocity field in an infinitesimally thin slice around the Supergalactic plane $SGZ=0$. We highlight the major structure present in this plane. Here, we use a 120\Mpch{} deep cut of the 2MASS Redshift Survey for  reconstructing the peculiar velocity field.}
\end{figure}

In Fig.~\ref{fig:flows}, we represented the reconstructed velocity
field in an infinitesimally thin cut around the Supergalactic plane. We
highlight the known large structures of the Local Universe with red
circles along with their names. It happens that all these major
structures lie on this plane. We also note the presence of large flows
falling towards this structures as expected. At the same time, there
is no divergence of these flows in the immediate neighborhood,
highlighting the quality of the MAK reconstruction.

 Now, we also note the presence of a relatively large void in the
 lower-left part of the panel. From this region the matter seems to
 escape. This pushes the idea that we not only study clusters of
 galaxies but also voids using the peculiar velocities.

\section{Studying void dynamics}
\label{sec:void_dyn}

In the previous section, we described a way to study peculiar velocity
dynamic using the MAK reconstruction method. We are now presenting how
this technique can be transported to the study of voids in large and
deep galaxy surveys.

\subsection{Identifying voids in large-scale structures}

\begin{figure}
  \includegraphics[width=\hsize]{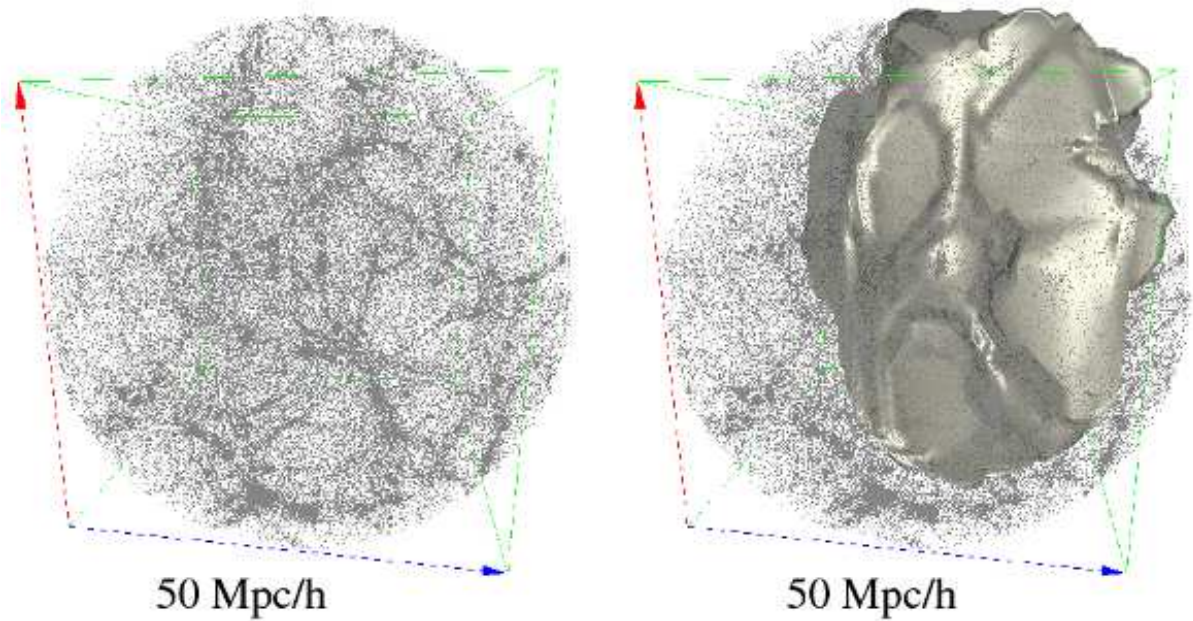}
  \caption{\label{fig:voids} {\it Voids in large-scale structures} --  Right panel: An example of a void found and whose volume has been determined in Lagrangian coordinates and transported to Eulerian coordinates (4\Mpch{} smoothing scale in Lagrangian coordinates). The dots correspond to the actual position of the particles of the simulation. We see that the highlighted boundary follows the distribution. The scale of the box is 50\Mpch{}. }
\end{figure}

Voids are large strongly underdense region of the Universe, that we know most probably forms by gravitational instabilities from primordial density fluctuations \citep{HoffmanShaham82,HSW83,HOR83}. One of these voids is illustrated in the particle distribution given by a $N$-body simulation in the left panel of Fig.~\ref{fig:voids}.

\begin{figure}
  \includegraphics[width=.7\hsize]{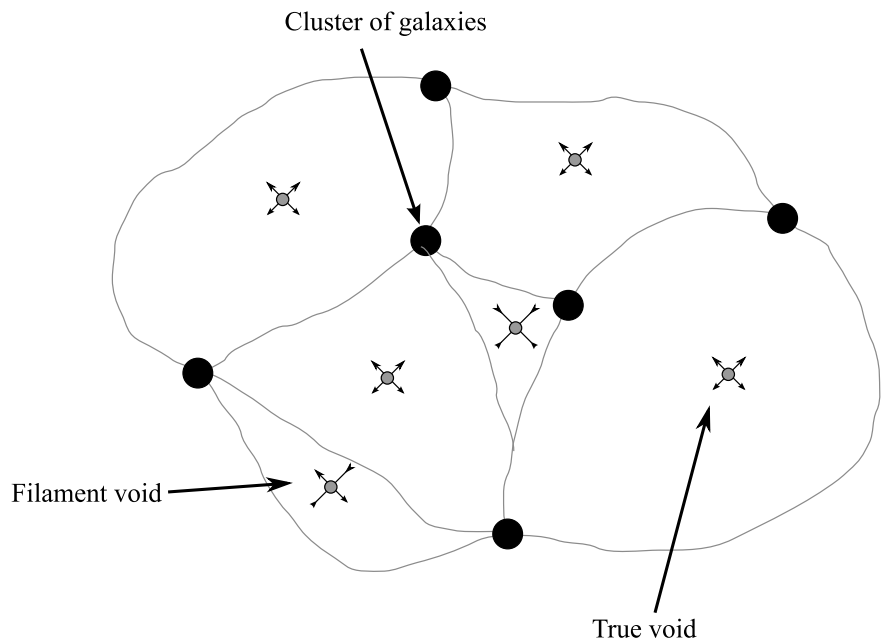}
  \caption{\label{fig:void_scheme} We give an illustration of the dynamical scheme we propose to study voids and why it is important. Actual observed voids may be contracting (represented by the arrows pointing towards the center of void)  or expanding (arrows going away from center). Contracting voids are not distinguished from expanding voids with standard Eulerian void finder techniques.}
\end{figure}

Since their discovery by \cite{GT78,TullyFisher78,Kirshner81} in
galaxy redshift surveys, a lot of work has gone into finding a
reliable way of detecting and characterizing the voids in large galaxy
catalogs. We can separate all these void finders in three broad
classes: galaxy based void finders, for which we look for holes in a
distribution of galaxies, density based void finders, for which we try
to find geometrical patterns in the total matter density field
inferred from galaxy distribution, and dynamical void finders, for
which gravitationally unstable points are derived from this same
distribution. 

Even though the number of void finders was flourishing, the large
physical size of the voids has for a long time hindered their use as a
probe for fundamental cosmological physics. With nowadays deep
photometric and spectroscopic surveys of galaxies, this starts being
possible to accumulate sufficient statistics. However, we still miss a
simple and usable definition of a void that would allow us to compute
statistical properties of these objects. Their potential of information is important as
they represent the counterpart of the cluster: we can count them,
describe their shape and their size, just as for clusters, which already
yields important constraints on cosmological parameters \citep[e.g.][ for the latest attempt]{Vikhlinin09}

We propose a new definition and a new technique for identifying voids in the matter density field, which is described in more details in \cite{LW09}. This technique 
is built upon the success of the
Monge-Amp\`ere-Kantorovitch reconstruction. We identify voids with the main source of
displacement field, such as illustrated in Fig.~\ref{fig:void_scheme},
which is properly reconstructed by the MAK method, after smoothing in
Lagrangian coordinates. This smoothing step means that we probe
structures in a particular dynamical collapse time. As voids are in
different collapse time, it is necessary to smooth at different scales
to probe all possible void statistics. Here, we will present the
results only for one smoothing scale. We choose the center of the
voids to correspond with the maximum of the divergence of the
displacement field, which is then considered to be sourced by the voids. The geometry of the voids
themselves may be determined subsequently by considering a watershed \citep{wvf}
transform of the divergence of the displacement field. The boundaries
are then transported using the displacement field to their proper
Eulerian position. An illustration of a void found using this
technique is given in the right panel of Fig.~\ref{fig:voids}. 

Using the first derivatives of the displacement at the position of the
center of the voids, we define the ellipticity, as in
\cite{ParkLee07}:
\begin{equation}
  \varepsilon = 1 - \sqrt{\frac{1 + \lambda_3}{1+\lambda_1}}
\end{equation}
with $\lambda_1 > \lambda_2 > \lambda_3$ the eigenvalues of the
Jacobian matrix of the displacement field. This ellipticity measures
the amount of local shear of the matter density field at the center of
the voids. On one hand, this ellipticity is linked to the overall ellipticity
of the volume, although in a non-trivial manner. On the other hand, being linked to a
reasonably linear quantity as the displacement field, the tidal ellipticity may be properly
modelled analytically and at same time measured in simulation and observational
data.

\subsection{Measuring Dark Energy properties}

We modeled the ellipticity statistics using Zel'dovich approximation and Gaussian random field statistics in \cite{LW09}. Compared to the work by \cite{ParkLee07}, we introduced correlations on the density/gravity curvature, as we essentially identify voids with the proto-voids in the primordial density field.

\begin{figure}
  \includegraphics[width=\hsize]{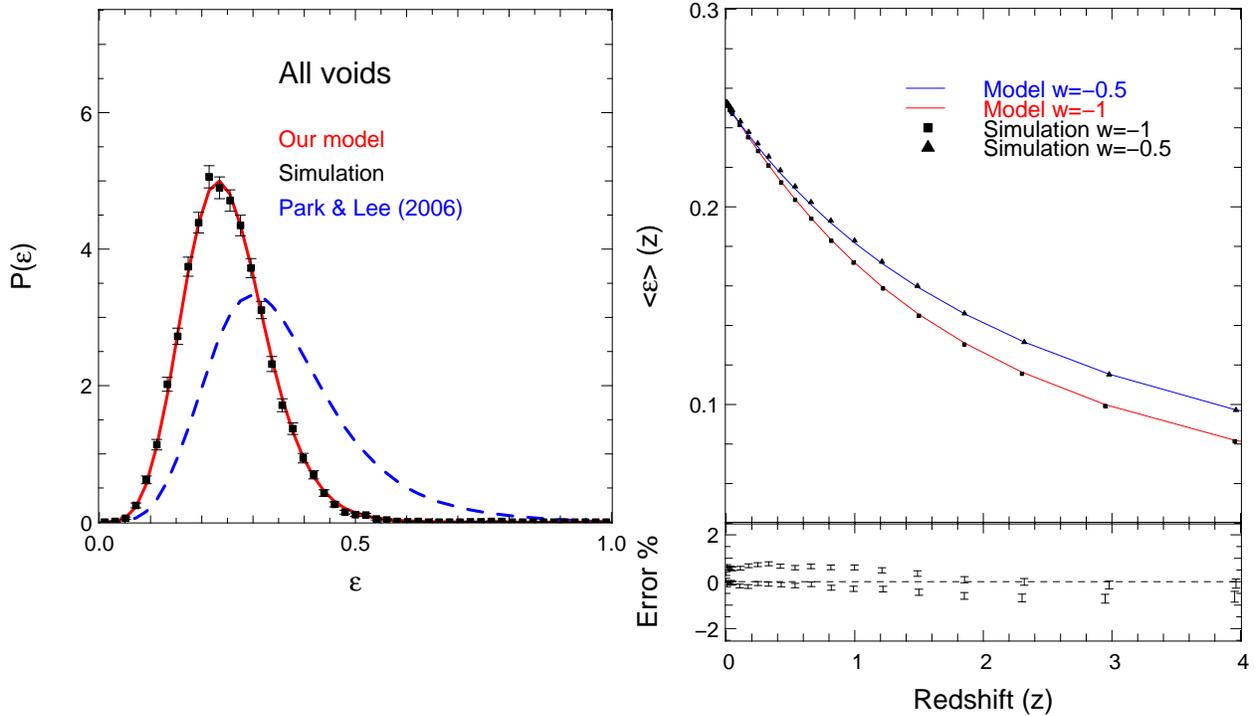}
  \caption{ \label{fig:epsilon_stat} {\it Ellipticity statistic modeling} -- Left panel: Comparison between the measured ellipticity distribution in simulation and the prediction given by our model and the original statistical model of \cite{ParkLee07}. The difference comes from the introduction of the correlation between density and gravity. Error bars on the points coming from simulation are estimated by assuming Poisson distribution on the binned values.  Right panel: Evolution of the mean ellipticity with redshift for the two $w$CDM cosmology with the residual between the actual measured points in the simulation and the model. The error bars on the residual are estimated from the intrinsic error in estimating the ellipticity and the number of averaged points. }
\end{figure}

In the left panel of Fig.~\ref{fig:epsilon_stat}, we show that the comparison between
analytic modeling (colored solid curves) and ellipticity measured in
simulation is impressive for our model. 

In the right panel of Fig.~\ref{fig:epsilon_stat}, we compare the
evolution of the mean ellipticity in two $w$CDM cosmologies, $w=-1$
and $w=-0.5$. The residual between the model and the actual
measurements in the simulation are shown in the bottom part of the
panel. The residual in not bigger than $\sim 1\%$ and most of it comes
from the actual variance of the initial condition of the
simulation. This effect will not prevent applying this method to
observations for two reasons. First, we will marginalize over the bias
and so the systematic shift will disappear. Second, each considered
slice should be a nearly independent random realization of a Gaussian
random field normalized to the same $\sigma_8$. Thus the points should
be scattered according to our dashed horizontal line ``0\%'' and not
systematically pushed up or down. We see that the points fall right
onto the curves given by the two models. This shows that this method
is a very promising tool to investigate dark energy physical property
like its equation of state $w$.

\section{Conclusion}
\label{sec:conclu}

In this paper, we have considered one particular method of reconstruction of peculiar velocities and applied it both to simulation and observation. Its performance at reconstructing simulated velocity field is impressive and far better than linear theory. This performance is now limited by our assumptions on the distribution of dark matter (the galaxy bias problem). 

We have applied our method with success on the Two-Micron-All-Sky redshift survey. The comparison between observed and reconstructed peculiar velocities is good and yielded an $\Omega_\text{m}$ in agreement with the value given by other cosmological probes, such as WMAP results \citep{WMAP5_LCDM}. In the future, we will aim at improving the statistics of the comparison.

It is possible to use reconstructed peculiar velocities for another related study: the identification of voids in galaxy surveys and their use for probing cosmological parameters such as the Dark Energy equation of state. As the displacement field, and the related peculiar velocity field, is still essentially linear, it is possible to make a simple model linking the local shape of voids, which is the ellipticity $\varepsilon$, to the background cosmology. We achieved this comparison with total success in $N$-body simulation. We will aim at using this method in real galaxy survey, like the Sloan Digital Sky Survey, to improve the constraints on $w$.

\begin{theacknowledgments}

I acknowledge financial support from NSF Grant AST 07-8849. This research was supported in part by the National Science Foundation through TeraGrid resources provided by the NCSA. Teragrid systems are hosted by Indiana University, LONI, NCAR, NCSA, NICS, ORNL, PSC, Purdue University, SDSC, TACC and UC/ANL. 

\end{theacknowledgments}

\bibliographystyle{aipproc}

%\bibliography{proceeding}

\end{document}